# A Timeout-Based Congestion Control Scheme for Window Flow-Controlled Networks

RAJ JAIN, SENIOR MEMBER, IEEE

*Abstract*—During overload, most networks drop packets due to buffer unavailability. The resulting timeouts at the source provide an implicit mechanism to convey congestion signals from the network to the source. On a timeout, a source should not only retransmit the lost packet, but it should also reduce its load on the network. Based on this realization, we have developed a simple congestion control scheme using the acknowledgment timeouts as indications of packet loss and congestion. This scheme does not require any new message formats, therefore, it can be used in any network with window flow control, e.g., ARPAnet or ISO.

## INTRODUCTION

A network with a high packet loss rate is perceived by its user as unreliable even if all hardware and software is up. Packets are lost if they cannot find buffers at intermediate nodes or at the destination. The problem of finding buffers at the destination is easily handled by appropriate flow control. The problem of ensuring availability of buffers at the intermediate nodes is rather difficult. In connectionless networks, which do not have a preset path for packets and do not have a predetermined set of intermediate nodes, the problem of ensuring buffers at the intermediate nodes is even more complicated than that in connection-oriented networks.

Generally, networks are highly underloaded, therefore, buffer availability is normally not an issue. However, during overload conditions, queues build up and the performance degrades severely due to buffer unavailability and the resulting loss of packets. The throughput may drop to zero and the response time may approach infinity during overload.

A strategy to reduce the impact of overload in a network is called *congestion control*. We distinguish between the terms *flow control* and *congestion control* as follows. *Flow control* is an agreement between a source and a destination to limit the flow of packets without taking into account the load on the network. The purpose of flow control is to ensure that a packet arriving at a destination will find a buffer there. Congestion control is primarily concerned with controlling the traffic to reduce overload on the network. Flow control limits traffic based on buffer availability at the destination, whereas, congestion control limits traffic based on buffer availability at intermediate nodes. Flow control is a bipartisan agree-ment. *Congestion control* is a social (network-wide) *law*. Different connections on a network can choose different flow control strategies but nodes on the network should follow the same congestion control strategy, if it is to be useful. It should be noted that there is considerable disagreement among researchers regarding the relationship between flow and congestion control. Some authors [2] consider congestion control to be a special case of flow control, while others [12] distinguish them as above.

A number of papers on flow and congestion control have been published. See [2] for an excellent survey. The key strategies used for dealing with congestion are the following [12]:

1) preallocation of resources to avoid congestion,

2) allowing intermediate nodes to discard packets at will,

3) restricting the number of packets allowed in the subnet,

4) using flow control, and

5) choking off input when congestion occurs.

In connectionless networks, such as ARPAnet and digital network architecture (DNA) paths for packets in a connection are not preset. Therefore, preallocation of resources at intermediate nodes (which are not known) is not feasible. Isarithmic schemes, which restrict the number of packets allowed in the subnet are difficult to implement and do not guarantee that a particular node will not get congested. Choking schemes generally require sending choke packets to sources of traffic when congestion occurs. Introducing additional packets in the networks during congestion is not desirable. Dropping packets without reducing the input to the network is not sufficient and may result in continuous packet losses leading to zero throughput.

Having discarded strategies 1 and 3 above, we propose to combine strategies 2, 4 and 5 in such a manner as to avoid their disadvantages. Actually, what we need is a mechanism for the network to tell the source that the network is congested and a mechanism for the source to adjust its load on the network. In window flow controlled networks, the window size provides an obvious way to adjust the load put by a source. Also since most networks drop packets during congestion due to buffer unavailability, the resulting timeouts at the source provide an implicit mechanism to convey congestion information from the network to the source. Explicit choke packets are not required.

Based on the above arguments we have developed a simple congestion control scheme called CUTE (congestion control using timeouts at the end-to-end layer). This scheme requires sources to reduce the load put on the network when congestion is sensed. The acknowledgment timeouts are used as indicators of packet loss and congestion.

This paper is organized as follows. It begins with a description of the network architecture assumed and the relationship between load on the network and flow control window sizes. We then describe the CUTE scheme, discuss alternative designs that were considered, and present performance graphs. The benefits of the proposed scheme in terms of ease of implementation are also discussed. More details on design and development of the CUTE scheme can be found in [7].

## ARCHITECTURAL ASSUMPTIONS

In designing the congestion control scheme, a number of assumptions have to be made about the underlying network architecture. For obvious reasons, we assumed the network to be a connectionless network with window flow control, which is what digitals network architecture (DNA) is. Although we used DNA as our network model, we believe that the final scheme can be used by any connectionless network that has window flow control, e.g., ARPAnet or ISO.

In window flow controlled networks, the key parameter determining the load on the network and hence its performance is the window sizes used by the sources. As the user increases the window size, throughput increases initially. The user is thus tempted to inject more and more packets into the network. However, if the window size is larger than a certain amount (depending upon the number of buffers at the intermediate nodes), the throughput starts dropping due to packet loss. The window size at the point of congestion is called the *buffer capacity* of the path. In the case of multiple users, the sum of the window sizes of all users passing through an intermediate node should be less than the buffer capacity of the node. Notice that the number of buffers at the bottleneck router generally determines the buffer capacity of a path.

## THE CUTE SSCHEME

The CUTE scheme requires sources to follow a set of self-restraining rules, so that the number of packets injected into the network by a source is limited, if congestion is sensed. The source dynamically adjusts its window size (WS) between a specified minimum and maximum as is shown in Fig. 1. On a timeout (packet loss), the window is reset to one. There are five specifications: a maximum, a minimum, an initialization policy, an increase policy, and a decrease policy. A number of alternatives were considered for each of these five specifications. The alternatives are compared in the next section. The scheme, as finally designed, is described in this section.

*1) Maximum:* The window size (WS) should never be

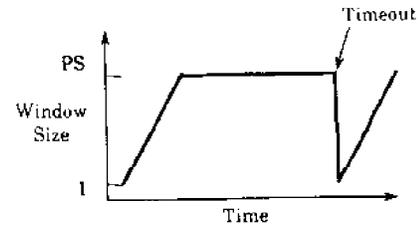

Fig. 1. Dynamic window adjustment using CUTE scheme.

more than a limit called *pipe size* (PS) which is specified by the network layer.

$$WS_{max} = PS$$

The network layer determines $PS$ from the currently known information about the path to the destination. Basically, $PS$ is the *optimal* number of packets that should be allowed on the path. Given line speeds and number of users, one can calculate $PS$. If line speeds are not available, we recommend that (for paths not including satellite links) $PS$ be three times the number of hops. For paths going through a satellite, $PS$ will need to be calculated using a more sophisticated algorithm using satellite bandwidth and propagation delay.

A source cannot send more packets than specified by the destination. Therefore,

$$WS_{max} = \min \{PS, \text{Credits from the destination}\}$$

If the number of hops is not available, then the only feasible alternative is to let the destination decide the maximum, i.e.,

$$WS_{max} = \text{Credits from the destination}$$

*2) Minimum:* The lower limit on $WS$ is one.

$$WS_{min} = 1$$

Thus, when reducing the window during congestion, the sources cannot reduce window size below one.

*3) Initialization:* The initial value of $WS$ is immaterial as long as it lies within the bounds specified by the above two rules. In a heavily loaded network it is safer to start at $WS_{min}$. In a lightly loaded network it is more efficient to start at $WS_{max}$. Other alternatives are to start at the $WS$ used during the last connection to the same node, or to start half-way between minimum and maximum.

In any case, the choice has little effect on the performance over a long period. The following adaptive features will soon bring $WS$ to a value suitable for the congestion state of the network.

*4) Increase:* $WS$ can be increased by one after the number of packets acknowledged since the last change (increase or decrease) becomes greater than or equal to the current value of $WS$. This gives a parabolic rise to $WS$ when plotted against packets acknowledged. Notice however, the rise is approximately linear in time because with $n$ packets outstanding, it takes one round-trip delay to get an acknowledgment for the $n$ packets. Thus, $WS$ increases by one every round-trip delay interval.

5) *Decrease:* On a timeout, the source should reset *WS* to the minimum allowed value.

$$WS \leftarrow WS_{min}$$

The next section is devoted to justification of these five rules and comparisons to other alternatives.

## ANALYSIS OF THE ALTERNATIVE DESIGNS

In order to study various alternatives, we developed a simulation model [6]. A simple configuration consisting of two local area networks (LAN's) interconnected via a slower speed wide area network (WAN) was simulated. Fig. 2 shows a logical representation of it. There are *n* sources sharing a common path through *m* intermediate nodes. There is no limitation on the number of sources or the number of intermediate nodes. Although most of the ideas came from the simulation, here we use simple analytical arguments to justify the rules of the proposed congestion control policy.

*1) Maximum:* The maximum determines the performance under no loss. Normally the window specified by the destination sets the upper limit. For destinations that allow very large windows, *PS* acts as the upper bound beyond which the performance gain is very little. Here, the performance is measured in terms of network power [3] which is defined as the ratio of total throughput to the average response time.

Given a path shared by *n* users, one can compute the optimal number of packets per user that should be allowed in the network. This number should be used as $WS_{max}$. It can be shown that for any terrestrial network, this number will always be less then $3h$, where *h* is the number of hops. This is therefore the recommended value in the absence of more information. The bound is arrived at as follows:

　　a) Using mean value analysis [11], it can be shown that a closed queueing network consisting of $N + 1$ *M*/*M*/1 queues, all with the same exponentially distributed mean service rate of *one*, and *C* customers circulating as shown in Fig. 3, will have:

$$\text{Throughput} = \frac{C}{N + C}$$

$$\text{Response time} = N + C$$

$$\text{Power} = \frac{\text{Throughput}}{\text{Response time}} = \frac{C}{(N + C)^2}$$

For Maximum Power, $\dfrac{dP}{dC} = 0 \Rightarrow C = N$

That is, the power is maximum when the number of customers is equal to the number of queues minus one. Using balanced job bounds [13], it can be shown that if server speeds are different, the number of customers *C* that optimizes power is even *less*. Increasing *C* beyond *N* *always* leads to a decrease in power.

　　b) An *h*-hop computer network can be represented by a closed queue in network as shown in Fig. 4. Packets to

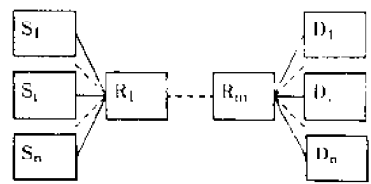

Fig. 2. A logical representation of the network configuration simulated.

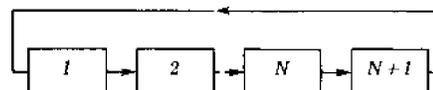

Fig. 3. A queueing network consisting of *C* customers circulating in $N + 1$ identical queues.

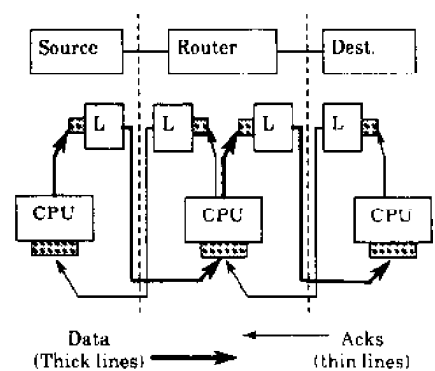

Fig. 4. An *h*-hop computer network contains $3h + 1$ queues.

be transmitted are queued by the source CPU to a transmit queue on the network layer, serviced at the line speed. These packets are then received at the router and put into a receive queue by the data-link layer. The CPU services this queue and puts the packets on an appropriate outgoing transmission queue. In this manner, the packet finally reaches the destination receive queue. The destination CPU generates an ack for the packet and the ack travels through various transmit queues and receive queues in the reverse direction. In Fig. 4, thick lines represent data packets and thin lines represent returning acks. The thinness signifies that the number of acks can be less than the number of packets, since some acks may acknowledge more than one packet. Also, the lines have been assumed to be full-duplex so that the traffic going in the forward direction is serviced by a different queue than that going in the reverse direction.

It is seen from the figure that, assuming full-duplex lines and one ack per packet, the number of queues in an h-hop network is $3h + 1$, consisting of:

| | |
|---|---|
| *h* | Forward transmit queues (*L*) used by data packets. |
| *h* | Reverse transmit queues (*L*) used by acks. |
| $h + 1$ | Receive queues serviced by node CPU's and shared by traffic in both directions. |

If the lines are half-duplex so that the traffic going in the two directions shares the same lines, the reverse transmit queues can be omitted and the number of queues would be $2h + 1$. Similarly, if the CPU's are so fast that the receive queues are serviced without queuing, $h$ CPU queues can be omitted from the model. Finally, if there are no acks, $h$ reverse transmit queues used by acks can be omitted. In this manner, given line types (full- or half-duplex), individual CPU speeds, line speeds, and average number of packets per ack, one can calculate the optimal $WS$. These variables vary from one network configuration to the next. In any case the optimal number will be between 1 and $3h$. There is never any gain in increasing window beyond $3h$ (excluding satellite networks). In [10], this analysis has been extended to networks with satellite links and it is shown that for such networks, the pipe size is approximately equal to the ratio of the minimum path delay (sum of all server's service times including propagation delay) to the bottleneck server's service time.

*2) Minimum:* For a minimum value of $WS$, we experimented with the following three alternatives:

a) $WS_{min} = 1$
b) $WS_{min}$ = number of hops
c) $WS_{min} = m$ times number of hops, where $m$ is a fixed parameter = 2, 3, 4, $\cdots$.

In our simulations, policy a) showed better performance than b) or c). The choice of the minimum impacts the upper limit on the number of users supportable. Since the total number of outstanding packets from all users should be below the buffer capacity of the path work, the best policy is to rise to the maximum as soon as possible, i.e., $N = 1$ is the best. For a highly congested network, the best policy is to keep the $WS$ at one and not rise at all, i.e., $N = \infty$ is the best. For intermediate levels of congestions, intermediate values of $N$ provide the best results.

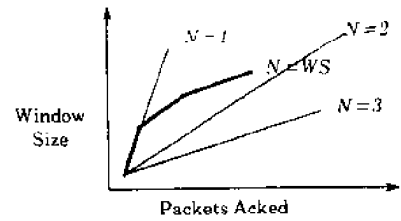

Fig. 5. Increase policies considered.

The parabolic increase policy eventually chosen provides a compromise for all situations. As the $WS$ rises, the chances for congestion increase, and the slope of rise decreases. Initially the scheme works like $N = 1$, and finally it works like $N = \infty$.

*5) Decrease Policy:* Among the five components of the CUTE scheme, the decrease policy turned out to be the most important in terms of its impact on the performance. We tried four different alternatives:

*a) Sudden Decrease: WS* is reduced to its minimum value.

$$number\ of\ connections * WS_{min} < buffer\ capacity\ of\ bottleneck\ router$$

$$max.\ number\ of\ connections < \frac{buffer\ capacity}{WS_{min}}$$

By choosing $WS_{min} = 1$, we are trying to support as many users as possible.

Another important reason for choosing $WS_{min} = 1$ is that it makes caching of out-of-order packets optional. In networks without out-of-order caching, loss of a single packet may lead to a cycle where every packet is transmitted twice. By allowing $WS$ to be reduced to one, it is possible to break this cycle. If we chose any higher value for $WS_{min}$, the cycle would continue indefinitely, degrading the performance seriously.

*3) Initialization:* This is probably the least important policy decision in the sense that network performance over a long period does not depend upon the initialization. The only case where the initial value matters is during short data transfers. In such cases, a good initial choice may slightly improve the performance. The CUTE scheme allows the implementors to choose any of the initialization policies discussed earlier.

*4) Increase Policy:* The first set of alternatives that we experimented with were a set of linear increase policies (see Fig. 5). The $N$th policy is to increment $WS$ by one for every $N$ successful transmissions. Thus, $N$ controls the slope of the rise as shown in Fig. 5. Larger $N$ results in slower rise.

The simulation showed that, for a lightly loaded net-

*b) Gradual Decrease: WS* is decreased by one on every timeout.

*c) Binary Decrease: WS* is reduced to half its value and rounded to an integer.

*d) Combinations of the above.*

The simulation results showed that in most cases, the sudden decrease policy results in the least loss and best performance. The explanation is as follows.

If the destination does not cache out-of-order packets, a single packet loss results in all subsequent arrivals being dropped at the destination. Thus, each real loss is followed by $n$ successive timeouts, where $n = WS$, and all of the above decrease policies will finally take $WS$ to one.

On the other hand, if the destination does cache out-of-order packets, and assuming that a timeout indicates a real overload, it can be shown that jumping to the minimum window size gives the least number of packet losses. Fig. 6 shows a hypothetical curve of $WS$ over time. Each peak represents a timeout and hence a packet loss. The binary decrease policy, for example, would result in twice as many packet losses as the sudden decrease policy. Similarly, other schemes would also result in higher losses.

Policies b)–d) show a performance superior to a) only if a significant number of timeouts are simply false alarms and not real overloads. One of the basic assumptions to be made in designing a policy is that the timeout parameters have been set correctly so that there are very few false alarms.

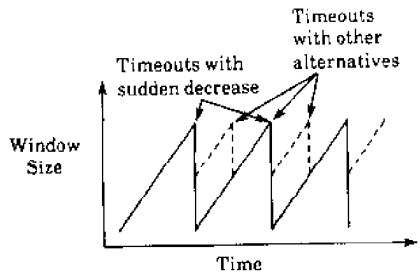

Fig. 6. If each timeout represents a real loss, a sudden decrease to the minimum window size gives the least number of losses as compared to any other alternative.

The proposed policy requires sources to reexplore the maximum after every timeout. An alternative scheme would be to remember the previously achieved maximum and not cross it. Such a scheme will lead to unfairness [7]. Also, every scheme with memory requires a *forgetting scheme* so that when the conditions improve the sources forget their previous maximum.

We have been using the CUTE policy since 1983 [4]. Since then, the fact that the best alternative on a timeout is to reduce the window size to one has also been discovered independently by Bux and Grillo [1] in the context of interconnection of their ring networks.

## PERFORMANCE

We compared the performance of various alternatives using metrics of throughput, response time (round-trip delay), power, fairness, and loss probability. For *n* users sharing the same path, the fairness is defined as follows:

$$\text{Fairness} = \frac{\left(\sum_{i=1}^{n} T_i\right)^2}{n \sum_{i=1}^{n} T_i^2}$$

where $T_i$ is the throughput of the *i*th user. The reasons for using this formula are discussed in [5].

In this section, we present only throughput curves. Other performance curves can be found in [7]. We simulated many different possible configurations. What follows are the cases that we believe provide meaningful insight.

*Case 1:* This is a case where the number of users sharing the same path is small but the congestion is caused by the destinations allowing large flow control windows. The window size permitted by the destination can be called *credits*. A source cannot increase its window size beyond the credits issued to it by the *destination* (CID). Fig. 7 shows performance as a function of CID for three sources sharing the same intermediate path. Each intermediate node has a buffer capacity of 42 packets. Depending on whether the source follows the CUTE policy, and whether the destination provides the *out-of-order* packets caching (OOC), there are four possibilities. Without dynamic window control, the throughput drops to zero as soon as the sum of credits exceeds the buffer capacity. The degradation is slower if there is out-of-order packet caching. With window control, the throughput is independent of credits after a certain maximum. Also, it does not matter (during this case of slight congestion) whether we have caching or not. This is because there are enough buffers to support $WS_{max}$ packets for each source and no packet is ever lost. The case of lost packets is discussed next.

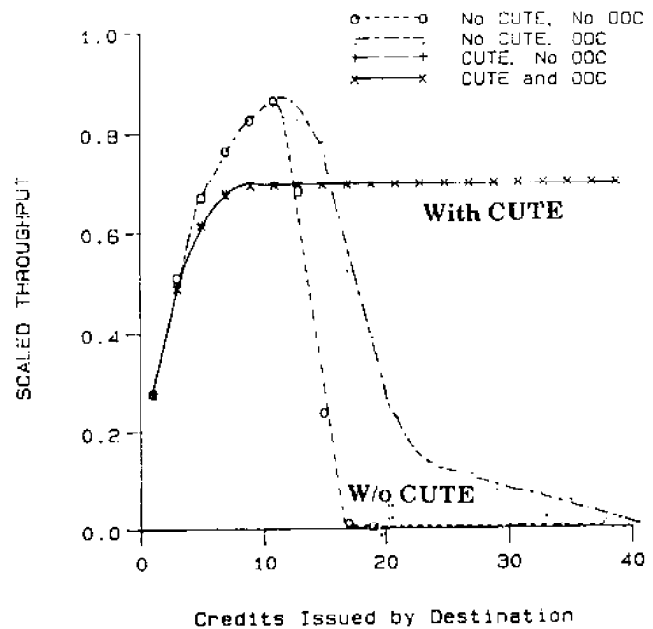

Fig. 7. Performance for Case 1.

*Case 2:* Fig. 8 shows network throughput as a function of number of sources sharing the same physical path for a network with moderate congestion. Each intermediate node has a buffer capacity of 80 packets, and each destination issues 8 credits. Without dynamic window control, the throughput drops to zero as soon as the sum of credits exceeds the buffer capacity. With window control, the total network throughput improves if there is out of order packet caching. Without caching, the throughput first drops, but eventually all sources operate effectively at single credit and caching does not matter.

## IMPLEMENTATION ISSUES

The following features of the CUTE scheme make it easy to implement in any computer network architecture.

1) It does not require changes to any message formats. No new bits are required.

2) It does not require any new messages. For example, no choke packets [9] are required.

3) It is transparent to network applications. Applications can issue any number of credits without worrying about number of buffers at intermediate nodes. The window stops increasing after reaching *pipe size* computed by the network layer.

4) Nodes with and without self-restraint can live on the same network. Nodes following the CUTE policy as well as those not following it lose packets during congestion. However, those not following the policy lose more packets than those following it. Nonetheless, as congestion is a network-wide phenomenon, it is preferable that all nodes follow the same policy.

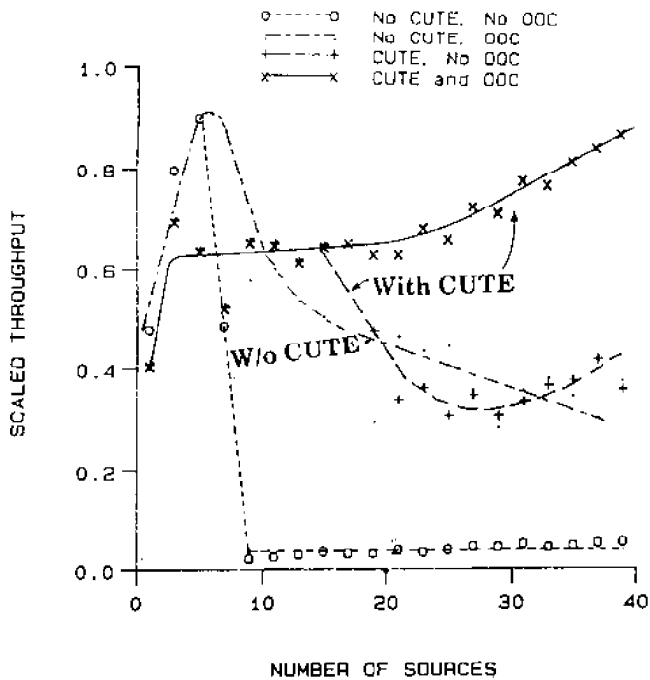

Fig. 8. Performance for Case 2.

5) The necessity of providing out-of-order caching is reduced. Without out-of-order caching, a single packet loss can result in all subsequent packets being transmitted twice. This cycle is broken by dropping the window size to one on a single loss. Thus, the performance without caching is at an acceptable level.

6) The scheme does not require any explicit feedback from the network. Black box schemes like CUTE are necessary if congestion (loss of packets) can occur at nodes that work transparently, e.g., bridges.

7) Window reduction at packet loss is necessary but not sufficient. More sophisticated congestion control schemes need to be developed that control the load on the network before it starts loosing packets. In networks having such mechanisms, CUTE can be used as a backup strategy which would be called only when congestion becomes so severe that it becomes necessary to drop packets.

## SUMMARY

The key ideas presented in this paper are: renewed importance of the congestion control, relationship between *timeout* and congestion, and a dynamic window control policy to handle congestion.

Solving the congestion problem has become more important due to the recent introduction of local area networks resulting in increased range of line speeds. Speed mismatch between incoming and outgoing lines results in queue buildup and packet loss.

The acknowledgment timeouts, which have traditionally been used to ensure end-to-end delivery, also indicate network congestion. On a timeout, therefore, we should take steps to reduce the load along with retransmitting the lost packet. If no steps are taken, the network may get into an infinite cycle of packet loss bringing the throughput down to zero.

Timeouts, in fact, indicate a *severe* state of congestion. As the congestion in the network increases, the queues build up. Only when the queueing becomes excessive, does it become necessary to drop packets. Once in this state, we need to take severe steps to reduce congestion. We should shut off all further transmissions until we are able to transmit successfully without loss. In particular, we should retransmit only one packet—the lost one. Also the window should be reduced to its smallest value. This leads us to the *key* rule in the proposed CUTE scheme: *on a timeout, reset the window to one.*

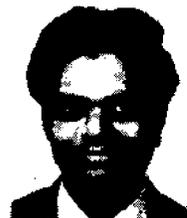

**Raj Jain** (S'73–M'78–SM'86) received the B.E. degree from A.P.S. University, Rewa, India, the M.E. degree from Indian Institute of Science, Bangalore, India, and the Ph.D. degree from Harvard University, Cambridge, MA, in 1972, 1974, and 1978, respectively.

His Ph.D. dissertation entitled "Control Theoretic Formulation of Operating Systems Resource Management Policies" was published by Garland Publishing, Inc. of New York in their "Outstanding Dissertations in the Computer Sciences" series. Since 1978, he has been with Digital Equipment Corporation, where he has been involved in performance modeling and analysis of a number of computer systems and networks including VAX Clusters, DECnet, and Ethernet. Currently, he is a Consulting Engineer in the Distributed Systems Architecture and Performance Group. He spent the 1983–1984 academic year on a sabbatical at the Massachusetts Institute of Technology doing research on the performance of networks and local area systems. For two years he has also taught a graduate course on computer system performance techniques at MIT and is writing a textbook on this subject.

Dr. Jain is a member of the Association for Computing Machinery.